\documentstyle[epsfig]{aipproc}

\begin{document}
\title{GRB Spectral Hardness \\
and Afterglow Properties}

\author{Jon Hakkila$^*$, Timothy Giblin$^{\dagger}$, \& Robert D. Preece$^{\dagger}$}
\address{$^*$Minnesota State University, Mankato, Minnesota 56001\\
$^{\dagger}$University of Alabama, Huntsville, Alabama 35812}

%\lefthead{LEFT head}
%\righthead{RIGHT head}
\maketitle

\begin{abstract}
A possible relationship between the presence of a radio afterglow and 
gamma-ray burst spectral hardness is discussed. The correlation is 
marginally significant; the spectral hardness of the bursts with 
radio afterglows apparently results from a combination of the break 
energy E$_{\rm break}$ and the high-energy spectral index $\beta$. 
If valid, this relationship would indicate that the afterglow does 
carry information pertaining to the GRB central engine.
\end{abstract}

\section*{Introduction}

Early observations of gamma-ray bursts (GRBs) with afterglows led 
us to hypothesize that GRBs with afterglows are spectrally harder 
than those without. However, the heterogeneous nature of GRB 
observations, coupled with the multi-wavelength nature of afterglow observations, has led us to a number of {\it concerns}:

\begin{enumerate}
\item Afterglow observations are heterogeneous: observational biases 
for detecting radio afterglows are different than those for detecting 
optical and/or x-ray afterglows.
\item A GRB might have no intrinsic afterglow, or inadequate search 
conditions might result in no afterglow being detected. 
\item Spectral hardness measures are biased by instrument performance: 
One instrumental dataset should be used.
\end{enumerate}

Our {\it solutions} to these concerns are as follows:

\begin{enumerate}
\item We require GRBs to either have detected radio afterglows or 
moderately-complete radio afterglow searches. This condition satisfies 
the first and second concerns.
\item We require that GRBs be observed by BATSE. This satisfies the third concern.
\end{enumerate}

Our resulting database is shown in Table \ref{table1}.
 
\begin{table}[ht!]
\caption{GRBs with and without radio afterglows.}

\label{table1}
%\begin{tabular}{lrrr}
\begin{tabular}{lcccccc}
%\begin{tabular}{lddd} 
GRB& Radio & HR(43/21)& HR21 &E$_{break}$& $\alpha$& $\beta$ \\
   & afterglow& & & (keV) \\
\tableline
%% \hline \\
970111&No&1.93 $\pm$ 0.03&1.38 $\pm$ 0.01&194.2 $\pm$ 4.0
&-1.14 $\pm$ 0.01&-3.65 $\pm$ 0.07 \\
970815&No&1.83 $\pm$ 0.32&0.93 $\pm$ 0.02&100.8 $\pm$ 57.9
&-0.82 $\pm$ 0.40&-3.09 $\pm$ 0.58 \\
971214&No&4.31 $\pm$ 0.34&1.67 $\pm$ 0.05&183.3 $\pm$ 36.7
&-1.14 $\pm$ 0.08&-2.85 $\pm$ 0.35 \\
980425&?&1.42 $\pm$ 0.38&0.96 $\pm$ 0.04&450.6 $\pm$ 766
&-1.67 $\pm$ 0.08&-11.0 $\pm$ 39.7 \\
970508&Yes&8.27 $\pm$ 1.32&1.09 $\pm$ 0.10&137.2 $\pm$ 96.9
&-1.18 $\pm$ 0.24&-1.88 $\pm$ 0.25 \\
970828&Yes&5.4 $\pm$ 0.2&1.81 $\pm$ 0.05&261.0 $\pm$ 44.2
&-0.09 $\pm$ 0.18&-3.72 $\pm$ 0.49 \\
980329&Yes&6.05 $\pm$ 0.09&1.31 $\pm$ 0.01&171.2 $\pm$ 4.78
&-1.28 $\pm$ 0.01&-2.22 $\pm$ 0.03 \\
980519&Yes&2.65 $\pm$ 0.33&1.05 $\pm$ 0.02&812 $\pm$ 1308
&-1.60 $\pm$ 0.03&-10.7 $\pm$ 37.3 \\
980703&Yes&6.36 $\pm$ 0.50&0.98 $\pm$ 0.03&181.8 $\pm$ 46.2
&-1.32 $\pm$ 0.06&-2.04 $\pm$ 0.13 \\
990123&Yes&18.0 $\pm$ 0.14&1.36 $\pm$ 0.01&618.6 $\pm$ 13.4
&-0.92 $\pm$ 0.01&-3.51 $\pm$ 0.28 \\
990506&Yes&5.15 $\pm$ 0.06&1.25 $\pm$ 0.01&258.7 $\pm$ 12.1
&-1.09 $\pm$ 0.03&-2.06 $\pm$ 0.05 \\
990510&Yes&1.87 $\pm$ 0.09&1.21 $\pm$ 0.02&120.8 $\pm$ 11.5
&-1.43 $\pm$ 0.04&-2.49 $\pm$ 0.07 \\
\end{tabular}
\end{table}

GRB 980425 may or may not have a radio afterglow, depending on its 
association with supernova SN 1998bw. If there is an association 
with SN 1998bw, then GRB 980425 has x-ray, optical, and radio afterglows. 
If there is no association with SN 1998bw, then this GRB has only an 
x-ray afterglow, with no optical or radio afterglow. 
The SAX team \cite{pian99} lists two possible x-ray afterglow sources 
for GRB 980425; the first one is consistent with SN 1998bw, the second 
one is not. 

The SAX Team indicates that the second afterglow source might have
``rebursted'', making its classification as a GRB counterpart 
questionable (this is not standard GRB behavior). However, it should
be noted that the flux measurement of the second observation of this 
source represents less than a $3 \sigma$ detection, placing the 
rebursting claim in doubt.

If the first afterglow source is associated with SN 1998bw, then 
GRB 980425 is significantly less luminous than typical GRBs. As 
discussed elsewhere in the literature \cite{schmidt99}, very few
BATSE GRBs can have luminosities this small in typical spatial 
distribution models.  

It appears to us that the afterglow source of GRB 980425 is still
in question. Because of these doubts, we consider independently the 
cases where GRB 980425 has and does not have a radio afterglow.

\section*{Are GRBs With Radio Afterglows Different from Those Without?}

\begin{figure}[ht!] % fig 1
\centerline{\epsfig{file=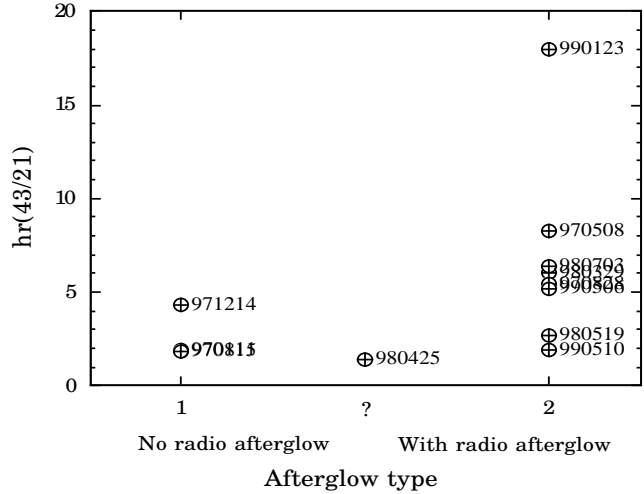,height=3in,width=3.5in}}
\vspace{10pt}
\caption{HR(43/21) vs. Radio Afterglow Type.}
\label{fig1}
\end{figure}

Figure \ref{fig1} demonstrates that GRBs with radio afterglows appear 
to be harder than those without. The hardness ratio HR(43/21) 
(100-1000 keV energy fluence divided by 25-100 keV energy fluence)
has been used in this analysis, because it has the largest 
signal-to-noise ratio available in BATSE 4-channel data and 
spans the largest spectral range. However, similar results
can be obtained from other hardness ratios.

In Table \ref{table2} we summarize Student's t-test probabilities 
that the $\log$[HR(43/21)] distributions of bursts without radio 
afterglows and bursts with radio afterglows have different means.
The significance of a correlation depends strongly on the status 
of GRB 980425 due to small number statistics. If GRB 980425 is 
not associated with SN 1998bw, then the correlation between spectral
hardness and radio afterglow is more likely. 

\begin{table}[ht!]
\caption{Probability that GRBs With and Without 
Radio Afterglows Have Different Distributional Means.}

\label{table2}
%\begin{tabular}{lrrr}
\begin{tabular}{lc}
%\begin{tabular}{lddd}
Status of SN 1998bw &t-Test Probability \\ 
\tableline
GRB980425 NOT associated with SN 1998bw &0.963 \\
GRB980425 associated with SN 1998bw &0.781 \\
\end{tabular}
\end{table}

\section*{Discussion}

\begin{figure}[ht!] % fig 2
\centerline{\epsfig{file=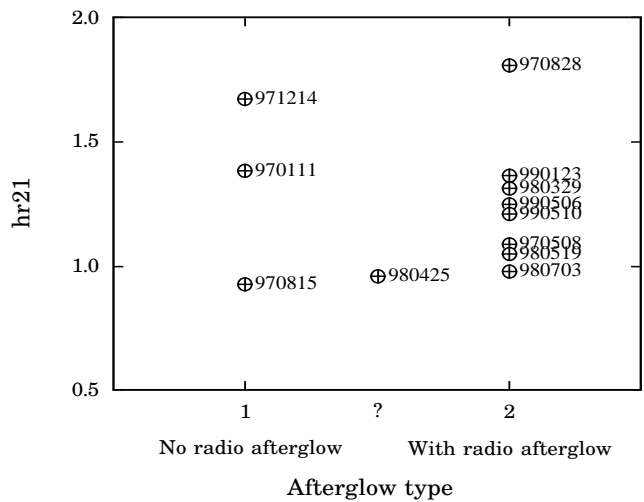,height=3in,width=3.5in}}
\vspace{10pt}
\caption{HR21 vs. Radio Afterglow Type.}
\label{fig2}
\end{figure}

To determine why this difference in hardness might exist, 
we checked other hardness ratios in the four-channel data. Hardness ratios 
involving channels 3 and 4 indicate similar results as obtained in 
Figure \ref{fig1}. Thus, any spectral differences dependent on radio 
afterglow type appear to result from the distribution of high energy 
photons. This is supported by Figure \ref{fig2}, which indicates no
correlation of radio afterglow type with hardness ratio HR21.

\begin{figure}[ht!] % fig 3
\centerline{\epsfig{file=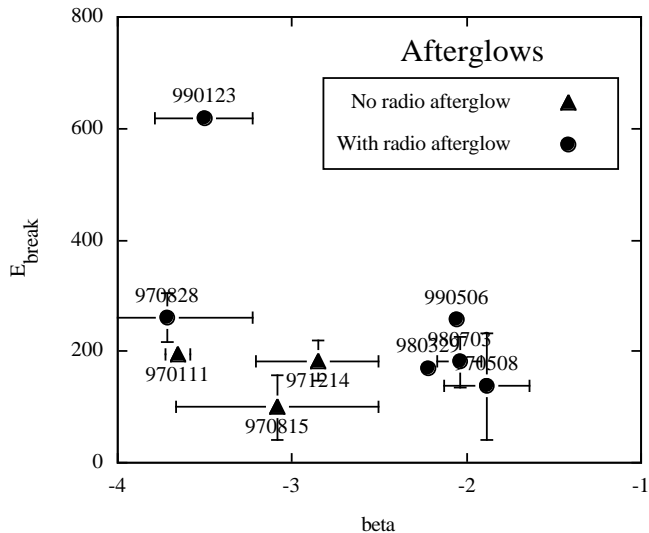,height=3in,width=3.5in}}
\vspace{10pt}
\caption{E$_{\rm break}$ vs. $\beta$ for different radio afterglow types.}
\label{fig3}
\end{figure}

This is also supported by Figure \ref{fig3}, which compares the GRB 
function spectral parameters E$_{\rm break}$ and $\beta$ for the 
bursts in question (GRBs 980425 and 980519 are not plotted due to 
large $\beta$ errors). 
Large values of E$_{\rm break}$, large values of $\beta$, or both 
produce conditions indicating many high-energy photons.  
GRBs with radio afterglows tend to occupy a different diagram region 
than GRBs without radio afterglows. Since E$_{\rm break}$ and $\beta$ 
are obtained from time-averaged spectra, we suspect that the diagram 
regions might be even more distinct if signal-to-noise were better for
faint BATSE GRBs. 

The possible correlation between GRB spectral hardness and 
afterglow type can be clarified with additional observations 
in the future. Resolution of the status of GRB 980425 would
also help clarify this issue. 

If GRB spectral hardness is an indicator of radio afterglow type,
then a direct link between the central engine and the delayed emission
is established. Such a link could be very important to the understanding
of GRB physics. The Lorentz factor of the expanding external shock could 
be constrained by this correlation. For this reason, it is as important
to determine upper flux limits on afterglow non-detections as it is to
provide information on detections.

\section*{Conclusions}

There is evidence that GRBs with radio afterglows have 
harder gamma-ray burst emission than those without. 
Due to small number statistics, the significance of this 
correlation depends at present time on whether or not GRB 980425 is 
associated with supernova SN 1998bw. 
There is evidence of a similar correlation between bursts with 
optical afterglows, but this is more difficult to document because the 
literature is less clear on conditions under which an optical search 
failed to yield an afterglow.

If we assume that a relationship exists between spectral hardness 
and radio afterglow type, then GRBs with radio afterglows appear to 
have more high-energy photons (E $\geq$ 100 keV) than those without 
radio afterglows, as determined from the spectral parameters 
E$_{\rm break}$  and $\beta$. Also, roughly $2/3$ of BATSE-detected 
GRBs should produce radio afterglows based on the overall distributions 
of E$_{\rm break}$ and $\beta$. It should be noted that all GRBs 
producing afterglows of any type belong to the long, bright, soft GRB 
class.

\section*{Acknowledgements}

We thank Dale Frail, Chip Meegan, Geoff Pendleton, Ralph Wijers, David 
Haglin, and Chryssa Kouveliotou for valuable discussions. Jon Hakkila 
acknowledges 1999 NASA/ASEE Summer Faculty Fellowship Program support.

\end{document}